\def\year{2022}\relax
\documentclass[letterpaper]{article} % DO NOT CHANGE THIS
\usepackage{aaai22}  % DO NOT CHANGE THIS
\usepackage{times}  % DO NOT CHANGE THIS
\usepackage{helvet}  % DO NOT CHANGE THIS
\usepackage{courier}  % DO NOT CHANGE THIS

\usepackage{graphicx} % DO NOT CHANGE THIS
\usepackage{natbib}  % DO NOT CHANGE THIS AND DO NOT ADD ANY OPTIONS TO IT
\usepackage{caption} % DO NOT CHANGE THIS AND DO NOT ADD ANY OPTIONS TO IT
\DeclareCaptionStyle{ruled}{labelfont=normalfont,labelsep=colon,strut=off} % DO NOT CHANGE THIS
\usepackage[vlined,boxed,commentsnumbered,linesnumbered,ruled]{algorithm2e}
\usepackage{amsmath,amscd,amsbsy,amssymb,amsfonts,latexsym,url,bm,amsthm,dsfont}
\usepackage{longtable}
\usepackage{booktabs}
\usepackage{algorithmic}
\usepackage{graphicx}
\usepackage{subfigure}
\usepackage{textcomp}
\usepackage{booktabs}
\usepackage{multirow}
\usepackage{graphicx}
\usepackage{threeparttable}
\usepackage{xcolor}
\usepackage{newfloat}
\usepackage{listings}
\title{Cross-Task Knowledge Distillation in Multi-Task Recommendation}
\author {
    Chenxiao Yang\textsuperscript{\rm 1},
    Junwei Pan\textsuperscript{\rm 2},
    Xiaofeng Gao\textsuperscript{\rm 1},
    Tingyu Jiang\textsuperscript{\rm 2},
    Dapeng Liu\textsuperscript{\rm 2},
    Guihai Chen\textsuperscript{\rm 1}
}
\affiliations {
    \textsuperscript{\rm 1} Department of Computer Science and Engineering, Shanghai Jiao Tong University\\
    \textsuperscript{\rm 2} Tencent Inc.\\
    chr26195@sjtu.edu.com, jonaspan@tencent.com, gao-xf@cs.sjtu.edu.cn, travisjiang@tencent.com, rocliu@tencent.com, gchen@cs.sjtu.edu.cn
}

\usepackage{bibentry}
\begin{document}

\maketitle

\begin{abstract}
% Multi-task learning (MTL) has been widely used in recommender systems to predict different types of user feedback (e.g, click, purchase). Most prior works focus on the design of network architectures for share-bottom layers in order to learn better representations of input features. However, the task-specific knowledge about users’ preferences towards different items, which is crucial for satisfactory ranking performance, has been overlooked under current MTL paradigm. To transfer such knowledge across tasks, this paper proposes a Cross-Task Knowledge Distillation framework: 1) \emph{Task Augmentation}: We introduce auxiliary tasks with quadruplet loss functions to capture cross-task fine-grained ranking information, which could avoid task conflicts by only preserving the task-consistent knowledge; 2) \emph{Knowledge Distillation}: We design a distillation approach to calibrate and distill ranking knowledge from augmented tasks to target tasks; 3) \emph{Model Training}: Teacher and student models are trained in an end-to-end manner, with a novel error correction mechanism to speed up model training and improve knowledge quality. Comprehensive experiments on a public dataset and our production dataset are carried out to verify the effectiveness of our framework as well as the necessity of its key components. 
Multi-task learning (MTL) has been widely used in recommender systems, wherein predicting each type of user feedback on items (e.g, click, purchase) are treated as individual tasks and jointly trained with a unified model. Our key observation is that the prediction results of each task may contain task-specific knowledge about user’s fine-grained preference towards items. While such knowledge could be transferred to benefit other tasks, it is being overlooked under the current MTL paradigm. This paper, instead, proposes a Cross-Task Knowledge Distillation framework that attempts to leverage prediction results of one task as supervised signals to teach another task. However, integrating MTL and KD in a proper manner is non-trivial due to several challenges including task conflicts, inconsistent magnitude and requirement of synchronous optimization. As countermeasures, we 1) introduce auxiliary tasks with quadruplet loss functions to capture cross-task fine-grained ranking information and avoid task conflicts, 2) design a calibrated distillation approach to align and distill knowledge from auxiliary tasks, and 3) propose a novel error correction mechanism to enable and facilitate synchronous training of teacher and student models. Comprehensive experiments are conducted to verify the effectiveness of our framework in real-world datasets.
\end{abstract}

\section{Introduction}
Online recommender systems often involve predicting various types of user feedback such as clicking and purchasing. Multi-Task Learning (MTL)~\cite{caruana1997multitask} has emerged in this context as a powerful tool to explore the connection of tasks for improving user interest modeling ~\cite{ma2018entire,lu2018like,wang2018explainable}.

\begin{figure}[t]
	\centering
	\includegraphics[width=0.49\textwidth,angle=0]{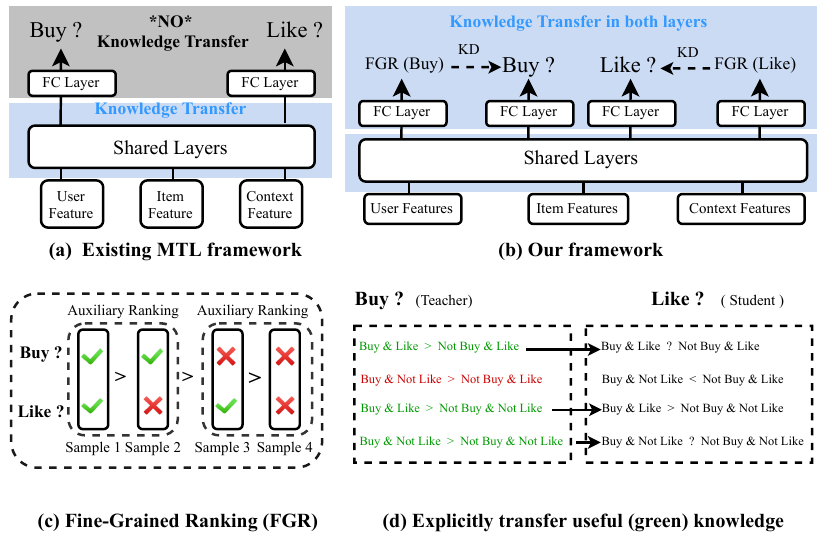}
	\caption{Illustration of the motivation of CrossDistil.}
	\label{fig_motiv}
\end{figure}

Common MTL models consist of low-level \emph{shared network} and several high-level \emph{individual networks}, as shown in Fig.~\ref{fig_motiv}(a), in the hope that the shared network could transfer the knowledge about ``how to encode the input features" by sharing or enforcing similarity on parameters of different tasks~\cite{ruder2017overview}. Most prior works~\cite{ma2018modeling,tang2020progressive,ma2019snr} put efforts on designing different shared network architectures with ad-hoc parameter-sharing mechanisms such as branching and gating.
In these models, each task is trained under the supervision of its own binary ground-truth label ($1$ or $0$), attempting to rank positive items above negative ones. However, using such binary labels as training signals, the task may fail to accurately capture user's preference for \emph{items with the same label}, despite that learning the auxiliary knowledge about these items' relations may benefit the overall ranking performance.

To address this limitation, we observe that the predictions of other tasks may contain useful information about how to rank same-labeled items. For example, given two tasks predicting `Buy' and `Like', and two items labeled as `Buy:$0$, Like:$1$' and `Buy:$0$, Like:$0$', the task `Buy' may not accurately distinguish their relative ranking since both of their labels are $0$. In contrast, another task `Like' will identity the former item as positive with larger probability (e.g. $0.7$), the latter with smaller probability (e.g. $0.1$). Based on the fact that a user is more likely to purchase the item she likes~\footnote{The same applies to other types of user feedback, e.g., click, collect, forward.}, we could somehow take advantage of these predictions from other tasks as a means to transfer ranking knowledge.

Knowledge Distillation (KD)~\cite{hinton2015distilling} is a teacher-student learning framework where the student is trained using the predictions of the teacher. As revealed by theoretical analysis in previous studies~\cite{tang2020understanding,phuong2019towards}, the predictions of the teacher, also known as \emph{soft labels}, are usually seen as more informative training signals than binary \emph{hard labels}, since they could reflect `whether the sample is true positive (negative)'. On the perspective of backward gradient, KD can adaptively re-scale student model's training dynamics based on the values of soft labels. Specially, in the above example, we could incorporate predictions $0.7$ and $0.1$ in the training signals for task `Buy'. Consequently, the gradients w.r.t the sample labeled `Buy:$0$ \& Like:$0$' in the example will be larger, indicating it is a more confident negative sample. Through this process, the task `Buy' could hopefully give accurate rankings of same-labeled items.
Motivated by these above observations, we proceed to design a new knowledge transfer paradigm on the optimization level of MTL models by leveraging KD. It is non-trivial due to three critical and fundamental challenges:
\begin{itemize}
    \item \textbf{How to address the task conflict problem during distillation?}\quad Not all knowledge from other tasks is useful~\cite{yu2020gradient}. Specially, in online recommendation, the target task may believe that a user prefers item$_A$ since she bought item$_A$ instead of item$_B$, while another task may reversely presume she prefers item$_B$ since she puts it in the collection rather than item$_A$. Such conflicting ranking knowledge may be harmful for the target task and could empirically cause significant performance drop.
    \item \textbf{How to align the magnitude of predictions for different tasks?}\quad Distinct from vanilla KD where teacher and student models have the same prediction target, different tasks may have different magnitude of positive ratio. Directly using another task's predictions as training signals without alignment could mislead the target task to yield biased predictions~\cite{zhou2021rethinking}.
    \item \textbf{How to enhance training when teacher and student are synchronously optimized?}\quad The vanilla KD adopts asynchronous training where the teacher model is well-trained beforehand. However, MTL inherently requires synchronous training, where each task is jointly learned from scratch. This indicates the teacher may be poorly-trained and provide inaccurate or even erroneous training signals, causing slow convergence and local optima~\cite{wen2019preparing,xu2020privileged}. 
\end{itemize}

In this paper, we propose a novel framework named as Cross-Task Knowledge Distillation (CrossDistil). Different from prior MTL models where knowledge transfer is achieved by sharing representations in bottom layers, CrossDistil also facilitates transferring ranking knowledge on the top layers, as shown in Fig.~\ref{fig_motiv}(b). To solve the aforementioned challenges: \textbf{First}, we introduce augmented tasks to learn the knowledge of the ranking orders of four types of samples as shown in Fig.~\ref{fig_motiv}(c). New tasks are trained based on a quadruplet loss function, and could fundamentally avoid conflicts by only preserving the useful knowledge and discarding the harmful one, as shown in Fig.~\ref{fig_motiv}(d). \textbf{Second}, we consider a calibration process that is seamlessly integrated in the KD procedure to align predictions of different tasks, which is accompanied with a bi-level training algorithm to optimize parameters for prediction and calibration respectively. \textbf{Third}, teachers and students are trained in an end-to-end manner with a novel error correction mechanism to speed up model training and further enhance knowledge quality. We conduct comprehensive experiments on a large-scale public dataset and a real-world production dataset that is collected from our platform. The results demonstrate that CrossDistil achieves state-of-the-art performance. The ablation studies also thoroughly dissect the effectiveness of its modules.

\section{Preliminaries and Related Works}

\paragraph{Knowledge Distillation}~\cite{hinton2015distilling} is a teacher-student learning framework where the student model is trained by mimicking outputs of the teacher model. 
% The prediction of the teacher model is often referred as `soft label', which is believed to contain knowledge reflected by subtle differences of predicted values for different samples. 
For binary classification, the distillation loss function is formulated as
\begin{equation} \label{eqn_vnkd}
    \mathcal L^{KD} = CE(\sigma(r_T/\tau), \sigma(r_S/\tau)),
\end{equation}
where $CE(y,\hat y) = y\log(\hat y) + (1-y)\log(1-\hat y)$ is binary cross-entropy, $r_T$ and $r_S$ denote logits of the teacher and student model, and $\tau$ is the temperature hyper-parameter. Recent advances~\cite{tang2020understanding,yuan2020revisiting} show that KD performs instance-specific label smoothing regularization that re-scales backward gradient in logits space, and thus could hint to the student model about the confidence of the ground-truth, which explain the efficacy of KD for wider applications apart from traditional model compression~\cite{kim2021self,yuan2020revisiting}.

% Empirical results reveal that KD could accelerate the training process and bring up model generalization ability. It has been found useful across a wide range of applications such as image classification, neural language processing and recommender systems. 

% An original purpose of KD is to distill useful information from a cumbersome teacher model into a lightweight student model~\cite{tang2018ranking,zhu2020ensembled}. 
% Though a common belief is that only a strong teacher model could teach weaker students, it is challenged by recent studies~\cite{cho2019efficacy,tang2020understanding}. Some works~\cite{yuan2020revisiting,zhang2020self} interpret KD from the perspective of intelligent label smoothing to explain its success in other applications besides model compression. These works inspire us to leverage KD as an explicit knowledge transfer approach rather than conventional model compression.

% KD has been developed for various applications apart from model compression, e.g., intelligent label smoothing~\cite{yuan2020revisiting}, self-distillation~\cite{zhang2020self}.

Existing works in recommender systems adopt KD for its original purpose, i.e., distilling knowledge from a cumbersome teacher model into a lightweight student model targeting \emph{the same task}~\cite{tang2018ranking,xu2020privileged,zhu2020ensembled}. Distinct from theirs or other KD works in other fields, this paper leverages KD to transfer knowledge \emph{across different tasks}, which is non-trivial due to the aforementioned three major challenges.

% and calls for deep and fundamental understanding of how KD works and its relation with ranking tasks in recommendation.
% , facing the following unique challenges: 1) student and teacher models are essentially solving different tasks with potentially contradictory ranking orders cross tasks; 2) the teacher model is trained with a ranking-based loss function, whose outputs have different magnitude with the student model and thus could not be directly applied for KD.

\paragraph{Multi-Task Learning}~\cite{zhang2021survey} is a machine learning framework that learns task-invariant representations by a shared bottom network, and yields predictions for each individual task by task-specific networks. It has received increasing interests in recommender systems~\cite{ma2018entire,lu2018like,wang2018explainable,pan2019predicting} for modeling user interests by predicting different types of user feedback. 
% The Shared-Bottom architecture~\cite{caruana1997multitask} is a classic MTL model and known to suffer from task conflict problem~\cite{misra2016cross,sener2018multi}. 
A series of works seek for improvements by designing different shared network architectures, such as adding constraints on task-specific parameters~\cite{duong2015low,misra2016cross,yang2016deep} and separating shared and task-specific parameters~\cite{ma2018modeling,tang2020progressive,ma2019snr}. Different from theirs, we resort to KD to transfer ranking knowledge across tasks on top of task-specific networks. Notably, our model is a general framework and could be leveraged as an extension of off-the-shelf MTL models.

\section{Proposed Model}

\begin{figure*}[t] 
	\centering
	\includegraphics[width=0.85\textwidth,angle=0]{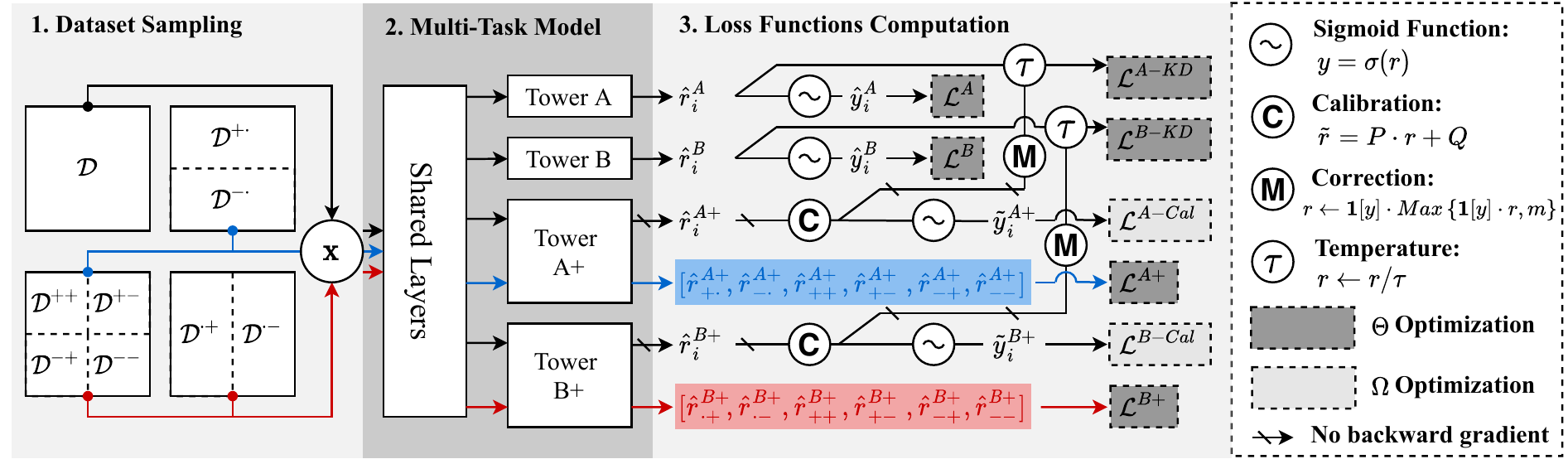}
	\caption{Illustration of computational graph for CrossDistil.}
	\label{fig_frame}
\end{figure*}

\subsection{Task Augmentation for Ranking} \label{sec_aug}
This paper focuses on multi-task learning for predicting different user feedback (e.g. click, like, purchase, look-through), and considers two tasks denoted as task $A$ and task $B$ to simplify illustration. As shown in the left panel of Fig.~\ref{fig_frame}, we first split the set of training samples into multiple subsets according to combinations of tasks' labels: 
\begin{equation}
\begin{split}
    % \mathcal D^{++} = \{\mathbf x_i \in \mathcal D &| y^A_i = 1, y^B_i = 1\}, \\ 
    \mathcal D^{+-} = \{(\mathbf x_i, y^A_i, y^B_i) &\in \mathcal D ~|~ y^A_i = 1, y^B_i = 0\},\\
    \mathcal D^{-+} = \{(\mathbf x_i, y^A_i, y^B_i) &\in \mathcal D ~|~ y^A_i = 0, y^B_i = 1\},\\
    % \mathcal D^{--} = \{\mathbf x_i \in \mathcal D &| y^A_i = 0, y^B_i = 0\},\\
    \mathcal D^{-\cdot} = \mathcal D^{--} \cup \mathcal D^{-+}&, \;\mathcal D^{+\cdot} = \mathcal D^{+-} \cup \mathcal D^{++},\\
    \mathcal D^{\cdot-} = \mathcal D^{--} \cup \mathcal D^{+-}&, \;\mathcal D^{\cdot+} = \mathcal D^{-+} \cup \mathcal D^{++},
\end{split}
\end{equation}
where $\mathbf x$ is an input feature vector, $y^A$ and $y^B$ denote hard labels for task $A$ and task $B$ respectively. The goal is to rank positive samples before negative ones, which can be expressed a bipartite order $\mathbf x_{_{+\cdot}} \succ \mathbf x_{_{-\cdot}}$ for task $A$ and $\mathbf x_{_{\cdot+}} \succ \mathbf x_{_{\cdot-}}$ for task $B$, where $\mathbf x_{_{+\cdot}} \in \mathcal D^{+\cdot}$ and so forth. 
Note that these bipartite orders may be contradictory among different tasks, e.g., $\mathbf x_{_{+-}} \succ \mathbf x_{_{-+}}$ for task $A$ while $\mathbf x_{_{+-}} \prec \mathbf x_{_{-+}}$ for task $B$. Due to the existence of such conflicts, directly conducting KD by treating one task as teacher and another task as student may cause inconsistent training signals and is empirically harmful for the overall ranking performance. 
% to backward gradients of shared parameters, leading to a negative affect on the overall prediction performance. Empirically,  by treating one task as the teacher and another task as the student fails to work due to these conflicts.

To enable knowledge transfer across tasks via KD, we introduce auxiliary ranking-based tasks that could essentially avoid task conflicts while preserving useful ranking knowledge. In specific, we consider a quadruplet $(\mathbf x_{_{++}}, \mathbf x_{_{+-}}, \mathbf x_{_{-+}}, \mathbf x_{_{--}})$ and the corresponding multipartite order $\mathbf x_{_{++}} \succ \mathbf x_{_{+-}} \succ \mathbf x_{_{-+}} \succ \mathbf x_{_{--}}$ for task A. In contrast with the original bipartite order, the multipartite order reveals additional information about the ranking of samples, i.e., $\mathbf x_{_{++}} \succ \mathbf x_{_{+-}}$ and $\mathbf x_{_{-+}} \succ \mathbf x_{_{--}}$ without introducing contradictions. Therefore, we refer such order as \emph{fine-grained ranking}. 
Based on this, we introduce a new ranking-based task called \emph{augmented task $A+$} for enhancing knowledge transfer by additionally maximizing
% the augmented tasks are designed to capture the underlying ranking relations of these samples. In specific, for augmented task A, the correct order of these samples is $\mathbf x_h \succ \mathbf x_i \succ \mathbf x_j \succ \mathbf x_k$. Note that such global ranking ensures the samples order is not contradictory to the original samples order, i.e., $\mathbf x_h$ or $\mathbf x_i \succ \mathbf x_j$ or $\mathbf x_k$. Similarly, the correct order for augmented task B is $\mathbf x_h \succ \mathbf x_j \succ \mathbf x_i \succ \mathbf x_k$. Following the inference process in ~\cite{rendle2012bpr}, the optimization target for augmented task A is
\begin{equation}\label{eqn-obj-a+}
\begin{split}
&\ln p(\Theta | \succ)\\
% = &\ln p(\mathbf x_{_{++}} \succ \mathbf x_{_{+-}} \wedge \mathbf x_{_{-+}} \succ \mathbf x_{_{--}} | \Theta) \cdot p(\Theta)\\
= &\ln p(\mathbf x_{_{++}} \succ \mathbf x_{_{+-}} | \Theta) \cdot p(\mathbf x_{_{-+}} \succ \mathbf x_{_{--}} | \Theta) \cdot p(\Theta)\\
= &\sum_{(\mathbf x_{_{++}}, \mathbf x_{_{+-}},\atop \mathbf x_{_{-+}}, \mathbf x_{_{--}})} \ln \sigma(\hat r_{_{++\succ+-}}) + \ln \sigma(\hat r_{_{-+\succ--}}) - Reg(\Theta),\\
\end{split}
\end{equation}
where $r$ is the logit value before activation in the last layer, $\hat r_{_{++\succ+-}} = \hat r_{_{++}} - \hat r_{_{+-}}$, and sigmoid function $\sigma(x) = 1/(1+exp(-x))$. The loss function for augmented task $A+$ is
\begin{equation} \label{eqn_apr}
\begin{split}
    \mathcal L^{A+} = &\sum_{(\mathbf x_{_{++}}, \mathbf x_{_{+-}}, \atop\mathbf x_{_{-+}}, \mathbf x_{_{--}})} -\beta_1^A \ln \sigma(\hat r_{_{++\succ+-}}) -\beta_2^A \ln \sigma(\hat r_{_{-+\succ--}})\\
    + &\sum_{(\mathbf x_{_{+\cdot}}, \mathbf x_{_{-\cdot}})} - \ln \sigma(\hat r_{_{+\cdot\succ-\cdot}}),
\end{split}
\end{equation}
which consists of three terms that respectively correspond to three pair-wise ranking relations of samples, where coefficients $\beta_1$, $\beta_2$ balance their importance. The loss function for augmented task $B+$ could be defined in a similar spirit.
% \begin{equation}
% \begin{split}
%     \mathcal L^{B+} &= \sum_{(\mathbf x_{_{++}}, \mathbf x_{_{+-}}, \atop \mathbf x_{_{-+}}, \mathbf x_{_{--}})} -\beta_1^B \ln \sigma(\hat r_{_{++\succ-+}}) -\beta_2^B \ln \sigma(\hat r_{_{+-\succ--}})\\
%     &+ \sum_{(\mathbf x_{_{\cdot+}}, \mathbf x_{_{\cdot-}})} - \ln \sigma(\hat r_{_{\cdot+\succ\cdot-}}).
% \end{split}
% \end{equation}
% The computational graph for augmented tasks are shown in highlighted in blue and red in Fig.~\ref{fig_frame}. 
These augmented ranking-based tasks are jointly trained with original regression-based tasks in MTL framework as shown in the second panel of Fig.~\ref{fig_frame}. The original regression-based loss function is formulated as:
\begin{equation} \label{eqn_la}
\begin{split}
    \mathcal L^{A} = &CE(y^A, \hat y^A), \qquad \mathcal L^{B} = CE(y^B, \hat y^B),\\
    CE(y,\hat y) &= \sum_{\mathbf x_i \in \mathcal D} - y_i \ln \hat y_i - (1- y_i) \ln (1-\hat y_i),
\end{split}
\end{equation}
where $\hat y = \sigma(r)$ is the predicted probability. 

The introduced auxiliary ranking-based tasks could avoid task conflicts and act as prerequisites for knowledge transfer through KD. Besides, the task augmentation approach itself is beneficial for the generalizability of main tasks~\cite{hsieh2021boosting} by introducing more related tasks that may provide hints about what shall be learned and transferred in shared layers.
% , verified by our empirical results

%  also show the auxiliary ranking tasks could help to improve the ranking performance of original tasks
% , presumably because they could provide hints about what shall be learned and transferred in shared layers.
% \cmt{, especially in the task-specific layers}.

% Different from the original tasks of predicting the probability of $\mathbf x_i$ being a positive feedback, the proposed augmented tasks focus on learning task-specific global order. Therefore, these tasks are naturally different and could serve as auxiliary learning objectives to benefit information sharing in the bottom layers.  

\subsection{Calibrated Knowledge Distillation} \label{sec_cal}
% A general idea of mainstream MTL frameworks~\cite{ma2018modeling,tang2020progressive,misra2016cross} is to improve knowledge transfer by designing complex task-sharing architectures. However, for one thing, such a generic paradigm that solely relies on task-invariant parameters would limit the ability of knowledge transfer. For another, since regression-based loss functions focus on fitting their own labels in one task, the high-level individual layers may over-learn the task-specific knowledge, overriding across-task knowledge transferred in low-level shared layers. 
We next design a cross-task knowledge distillation approach that can transfer fine-grained ranking knowledge for MTL. 
Since the prediction results of another task may contain the information about unseen rankings between samples of the same label, a straightforward approach is to use soft labels of another task to teach the current task by the vanilla hint loss (i.e. distillation loss) as in Eqn.~\eqref{eqn_vnkd}. Unfortunately, such naive approach may be problematic and even imposes negative effects in practice. This is because the labels of different tasks may have contradictory ranking information that would harm the learning of other tasks as mentioned previously. To avoid such conflicts, we instead treat augmented ranking-based tasks as teachers, original regression-based tasks as students, and adopt the following distillation loss functions:

\begin{equation} \label{eqn_dis}
\begin{split}
    \mathcal L^{A-KD} &= CE(\sigma(\hat r^{A+} / \tau), \sigma(\hat r^A/ \tau)),\\
    \mathcal L^{B-KD} &= CE(\sigma(\hat r^{B+}/ \tau), \sigma(\hat r^{B} / \tau)).
\end{split}
\end{equation}
Note that soft labels $\hat y^{A+} = \sigma(\hat r^{A+} / \tau)$ and $\hat y^{B+} = \sigma(\hat r^{B+} / \tau)$ are invariant when training the student model, and hence the student will not mislead the teacher. The loss functions for students are formulated as
\begin{equation} \label{eqn_stu}
\begin{split}
    \mathcal L^{A-Stu} &= (1 - \alpha^A) \mathcal L^{A} + \alpha^A \mathcal L^{A-KD},\\
    \mathcal L^{B-Stu} &= (1 - \alpha^B) \mathcal L^{B} + \alpha^B \mathcal L^{B-KD},\\
\end{split}
\end{equation}
where $\alpha^A \in [0,1]$ is the hyper-parameter to balance two losses. The soft labels output by augmented ranking-based tasks are more informative training signals than hard labels. As an example, for samples $\mathbf x_{_{++}}, \mathbf x_{_{+-}}, \mathbf x_{_{-+}}, \mathbf x_{_{--}}$, the teacher model for augmented task $A+$ may give predictions $0.9$, $0.8$, $0.2$, $0.1$ which intrinsically contains auxiliary ranking orders $\mathbf x_{_{++}} \succ \mathbf x_{_{+-}}$ and $\mathbf x_{_{-+}} \succ \mathbf x_{_{--}}$ that are not revealed in hard labels. Such knowledge is then explicitly transferred through the distillation loss and can meanwhile regularize task-specific layers from over-fitting the hard labels.

However, an issue of the aforementioned approach is that augmented tasks are optimized with pair-wise loss functions and thus are not predicting a probability, i.e., the prediction $\sigma(\hat r^{A+})$ does not agree with the actual probability that the input sample is a positive one. 
Directly using the soft labels of teachers may mislead students and cause performance deterioration. To solve this problem, we propose to calibrate the predictions so as to provide numerically sound and unbiased soft labels. Platt Scaling~\cite{niculescu2005predicting,platt1999probabilistic} is a classic probability calibration method. We adopt it for calibration in this work. Still, one can replace it with any other more complex methods in practice. Formally, to get calibrated probabilities, we transform the logit values of teacher models through the following equation:
% \begin{equation}
% \begin{split}
%     \tilde y^{A+} = \frac{1}{1+\exp{P^A\cdot \hat r^{A+} + Q^A}}, \; \tilde y^{B+} = \frac{1}{1+\exp{P^B\cdot \hat r^{B+} + Q^B}}
% \end{split}
% \end{equation}
\begin{equation}
\begin{split}
    \tilde r^{A+} = P^A \cdot \hat r^{A+} + Q^A, \;\; \tilde y^{A+} = \frac{1}{1+\exp{ \tilde r^{A+}}}
\end{split}
\end{equation}
where $\tilde r$ and $\tilde y$ are the logit value and probability after calibration, respectively. The same process is also used for task $B+$. $P$, $Q$ are learnable parameters specific to each task. They are trained by optimizing the calibration loss
\begin{equation} \label{eqn_cal}
    \mathcal L^{Cal} = \mathcal L^{A-Cal} + \mathcal L^{B-Cal} = CE(y^A, \tilde y^{A+}) + CE(y^B,\tilde y^{B+}).
\end{equation}
We fix MTL model parameters when optimizing $\mathcal L^{Cal}$ as shown in the third panel of Fig.~\ref{fig_frame}. Note that, since the calibrated outputs of the teacher model are linear projections of the original outputs, the ranking result is unaffected so that the latent fine-grained ranking knowledge in soft labels is preserved during the calibration process. Distillation losses in Eqn.~\eqref{eqn_dis} are then revised by replacing $\hat r^{A+}, \hat r^{B+}$ with $\tilde r^{A+}, \tilde r^{B+}$.

\subsection{Model Training} \label{sec_train}
Conventional KD adopts a two-stage training process where the teacher model is trained in advance and its parameters are fixed when training the student model~\cite{hinton2015distilling}. However, such asynchronous training procedure is not favorable for industrial applications such as online advertising. Instead, due to simplicity and easy maintenance, synchronous training procedure where teacher and student models are trained in an end-to-end manner is more desirable as done in~\cite{xu2020privileged,anil2018large,zhou2018rocket}. In our framework, there are two sets of parameters for optimization, namely, parameters in MTL backbone for prediction (denoted as $\Theta$) and parameters for calibration including $P^A$, $P^B$, $Q^A$ and $Q^B$ (denoted as $\Omega$). To jointly optimize prediction parameters and calibration parameters, we propose a bi-level training procedure where $\Theta$ and $\Omega$ are optimized in turn for each iteration as shown in the training algorithm. For sampling, it is impractical to enumerate every combination of samples as in Eqn.~\eqref{eqn_apr}. Instead, We adopt bootstrap sampling strategy as used in~\cite{rendle2012bpr,shan2018combined} as unbiased approximation. 

\begin{algorithm}[t]
\algsetup{linenosize=\small}
\small
\SetAlgoLined
\KwIn{Training dataset $\mathcal{D}$, learning rate $\gamma_1$ and $\gamma_2$, initial parameters $\Theta$ and $\Omega$.}
Construct set $\mathcal D^{++}, \mathcal D^{+-},\mathcal D^{-+},\mathcal D^{--}, \mathcal D^{+\cdot}, \mathcal D^{-\cdot},\mathcal D^{\cdot+},\mathcal D^{\cdot-}$\;
%  $t \gets 0$\;
 \While{Not converged}{
 Sample $\mathbf x$ uniformly at random from $\mathcal D$\;
 Sample $\mathbf x_{_{++}}, \mathbf x_{_{+-}}, \mathbf x_{_{-+}}, \mathbf x_{_{--}}$ uniformly at random from $\mathcal D^{++}, \mathcal D^{+-},\mathcal D^{-+},\mathcal D^{--}$ respectively\;
 Sample $\mathbf x_{_{+\cdot}}, \mathbf x_{_{-\cdot}}, \mathbf x_{_{\cdot+}}, \mathbf x_{_{\cdot-}}$ uniformly at random from $\mathcal D^{+\cdot}, \mathcal D^{-\cdot},\mathcal D^{\cdot+},\mathcal D^{\cdot-}$ respectively\;
 \textbf{Model parameter $\Theta$ optimization}:\\
 Calculate $\mathcal L^{A+}(\mathbf x_{_{+\cdot}}, \mathbf x_{_{-\cdot}}, \mathbf x_{_{++}}, \mathbf x_{_{+-}}, \mathbf x_{_{-+}}, \mathbf x_{_{--}}; \Theta)$\;
 Calculate $\mathcal L^{B+}(\mathbf x_{_{\cdot+}}, \mathbf x_{_{\cdot-}}, \mathbf x_{_{++}}, \mathbf x_{_{+-}}, \mathbf x_{_{-+}}, \mathbf x_{_{--}}; \Theta)$\;
 Calculate $\mathcal L^{A-Stu}(\mathbf x; \Theta), \mathcal L^{B-Stu}(\mathbf x; \Theta)$\;
$\mathcal L^{Model} \gets wightedSum( \mathcal L^{A+}, \mathcal L^{B+}, \mathcal L^{A-Stu}, \mathcal L^{B-Stu})$\;
$\Theta \gets \Theta - \gamma_1 \nabla_{\Theta} \mathcal L^{Model}$\;
 \textbf{Calibration parameter $\Omega$ optimization}:\\
 Calculate $\mathcal L^{Cal}(\mathbf x; \Omega)$\;
$\Omega \gets \Omega - \gamma_2 \nabla_{\Omega} \mathcal L^{Cal}$\;
 }
\caption{Training Algorithm for CrossDistil} \label{alg_train}
\end{algorithm}

\subsection{Error Correction Mechanism}
In KD-based methods, the student model is trained according to predictions of the teacher model, without considering if they are accurate or not. 
However, inaccurate predictions of the teacher model that is contradictory with the hard label could harm the student model's performance in two aspects. First, at early stage of training when the teacher model is not well-trained, frequent errors in soft labels may distract the training process of the student model, causing slow convergence~\cite{xu2020privileged}.
% and result in a worse trained student model. In turn, a worse student model harms the teacher model through shared parameters in the bottom layers. Such loop slows down the training process. 
Second, even at later stage of training when the teacher model is relatively well-trained, it is still likely that the teacher model would occasionally provide mistaken predictions that may cause performance deterioration~\cite{wen2019preparing}. A previous work~\cite{xu2020privileged} adopts a warm-up scheme by removing distillation loss in the earliest $k$ steps of training. However, it is not clear how to choose an appropriate hyper-parameter $k$, and it cannot prevent errors after $k$ steps. 

In this work, we propose to adjust predictions of the teacher model $\tilde y$ to align with the hard label $y$. Specifically, we clamp logit values for the teacher model (if the prediction is inconsistent with the ground truth) as follows:
\begin{equation} \label{eqn_corr}
     r^{Teacher}(\mathbf{x}) \leftarrow  \mathds{1}[y]\cdot Max\left\{\mathds{1}[y] \cdot r^{Teacher}(\mathbf{x}),m\right\}
\end{equation}
where $r^{Teacher}$ could be $\tilde r^{A+}$ or $\tilde r^{B+}$, $\mathds{1}[y]$ is an indicator function that returns $1$ is $y=1$ else returns $-1$, and $m$ is the error correction margin, a hyper-parameter.  This procedure could accelerate convergence by eliminating inaccurate predictions at the early stage of training, and further enhance knowledge quality at the later stage to improve student model's performance. 
The proposed error correction mechanism has the following properties: 1) It does not affect the predictions of the teacher model if they are sufficiently correct (that predicts the true label with at least probability $\sigma(m)$); 2) It does not affect training of the teacher model since the computation of distillation loss has no backward gradient for teachers as shown in Fig.~\ref{fig_frame}.

% In the training process, they are optimized in turn.  Algorithm~\ref{alg_train} shows the training procedure.

\section{Experiments}
We conduct experiments on real-world datasets to answer the following research questions: \textbf{RQ1}: How do CrossDistil performs compared with the state-of-the-art multi-task learning frameworks; \textbf{RQ2}: Are the proposed modules in CrossDistil effective for improving the performance; \textbf{RQ3}: Does error correction mechanism help to accelerate convergence and enhance knowledge quality; \textbf{RQ4}: Does the student model really benefit from auxiliary ranking knowledge; \textbf{RQ5}: How do the hyper-parameters influence the performance?

% \subsection{Experiment Settings}
\subsection{Datasets}
We conduct experiments on a publicly accessible dataset TikTok~\footnote{https://www.biendata.xyz/competition/icmechallenge2019/data/} and our WechatMoments dataset. 
Tiktok dataset is collected from a short-video app with two types of user feedback, i.e., `Finish watching' and `Like'. WechatMoments dataset is collected through sampling user logs during 5 consecutive days with two types of user feedback, i.e., `Not interested' and `Click'.
For Tiktok, we randomly choose 80\% samples as training set, 10\% as validation set and the rest as test set. For WechatMoments, we split the data according to days and use the data of the first four days for training and the last day for validation and test. The statistics of datasets are given in Table~\ref{tab:dataset}.

\begin{table}[h]
	\centering
	\normalsize
	\caption{Statistics of two datasets.}
% 	\vspace{-5pt}
	\label{tab:dataset}
	\setlength{\tabcolsep}{1mm}{
	\scalebox{0.80}{
	\begin{tabular}{c|c|c|c|c|c}
		\toprule
		Datasets & \#Samples & \#Fields & \#Features & Density(A) & Density(B) \\
		\midrule
		WechatMoments & 9,381,820 & 10 & 447,002 & 1.510\% & 9.975\% \\
		TikTok & 19,622,340 & 9 & 4,691,483 & 37.994\% & 1.101\%\\
		\bottomrule
	\end{tabular}}}
% 	\vspace{-10pt}
\end{table}

\begin{table*}[t]
	\centering
	\caption{Experiment results of CrossDistil and competitors on WechatMoments dataset.}
% 			\vspace{-5pt}
	\label{tab_result2}
	\scalebox{0.74}{
	\begin{tabular}{c|c|c|c|c|c|c|c|c}
		\toprule[1pt]
		\specialrule{0em}{1pt}{1pt}
		\multirow{2}{*}{Methods}& \multicolumn{2}{c|}{TaskA-Student}  & \multicolumn{2}{c|}{TaskB-Student}  & \multicolumn{2}{c|}{TaskA-Teacher}& \multicolumn{2}{c}{TaskB-Teacher} \\
		\specialrule{0em}{1pt}{1pt} \cline{2-9}
		\specialrule{0em}{1pt}{1pt}
		& AUC    & Multi-AUC    & AUC    & Multi-AUC  & AUC  & Multi-AUC& AUC  & Multi-AUC \\
		\specialrule{0em}{1pt}{1pt} \hline \specialrule{0em}{1pt}{1pt}
	    Single-Model  & 0.7528 & 0.6270 & 0.7597 & 0.6024 & 0.7535 & 0.6708 & 0.7604 & 0.6705 \\
	    Shared-Bottom & 0.7540$_{(+0.0012)}$ & 0.6378$_{(+0.0108)}$ & 0.7587$_{(-0.0010)}$ & 0.6145$_{(+0.0121)}$ & -&-&-&-\\
	    Cross-Stitch  & 0.7582$_{(+0.0054)}$ & 0.6360$_{(+0.0090)}$ & 0.7600$_{(+0.0003)}$ & 0.6195$_{(+0.0171)}$ & -&-&-&-\\
	    MMoE  & 0.7619$_{(+0.0091)}$ & \underline{0.6431}$_{(+0.0161)}$ & 0.7605$_{(+0.0008)}$ & 0.6226$_{(+0.0202)}$ & -&-&-&-\\
	    PLE &  \underline{0.7625}$_{(+0.0097)}$ & 0.6394$_{(+0.0124)}$ &\underline{0.7607}$_{(+0.0010)}$ & \underline{0.6240}$_{(+0.0216)}$ & -&-&-&-\\
	    \specialrule{0em}{1pt}{1pt} \hline \specialrule{0em}{1pt}{1pt}
	    TAUG  & 0.7632$_{(+0.0104)}$ & 0.6432$_{(+0.0162)}$ & 0.7612$_{(+0.0015)}$ & 0.6394$_{(+0.0370)}$& \textbf{0.7625}$_{(+0.0090)}$ & 0.6853$_{(+0.0145)}$ & 0.7608$_{(+0.0004)}$ & 0.6768$_{(+0.0063)}$ \\
	    CrossDistil  & \textbf{0.7644}$_{(+0.0116)}$ & \textbf{0.6879}$_{(+0.0609)}$ & \textbf{0.7618}$_{(+0.0021)}$ & \textbf{0.6861}$_{(+0.0837)}$  & 0.7618$_{(+0.0083)}$ & \textbf{0.6910}$_{(+0.0202)}$ & \textbf{0.7609}$_{(+0.0005)}$ & \textbf{0.6850}$_{(+0.0145)}$\\
		\specialrule{0em}{1pt}{1pt}
		\bottomrule[1pt]
	\end{tabular}}
% 	\vspace{-6pt}
\end{table*}

\begin{table*}[t]
	\centering
	\caption{Experiment results of CrossDistil and competitors on TikTok dataset.}
% 		\vspace{-5pt}
	\label{tab_result1}
	\scalebox{0.74}{
	\begin{tabular}{c|c|c|c|c|c|c|c|c}
		\toprule[1pt]
		\specialrule{0em}{1pt}{1pt}
		\multirow{2}{*}{Methods}& \multicolumn{2}{c|}{TaskA-Student}  & \multicolumn{2}{c|}{TaskB-Student}  & \multicolumn{2}{c|}{TaskA-Teacher}& \multicolumn{2}{c}{TaskB-Teacher} \\
		\specialrule{0em}{1pt}{1pt} \cline{2-9}
		\specialrule{0em}{1pt}{1pt}
		& AUC    & Multi-AUC    & AUC    & Multi-AUC  & AUC  & Multi-AUC& AUC  & Multi-AUC \\
		\specialrule{0em}{1pt}{1pt} \hline \specialrule{0em}{1pt}{1pt}
	    Single-Model & 0.7456 & 0.6335 & 0.9491 & 0.7966 & 0.7453 & 0.7140 & 0.9481 & 0.8297\\
	    Shared-Bottom & 0.7375$_{(-0.0081)}$ & 0.6344$_{(+0.0009)}$ & 0.9489$_{(-0.0002)}$ & \underline{0.8101}$_{(+0.0135)}$ & -&-&-&-\\
	    Cross-Stitch & 0.7468$_{(+0.0012)}$ & 0.6445$_{(+0.0110)}$ & 0.9488$_{(-0.0003)}$ & 0.7985$_{(+0.0019)}$ & -&-&-&-\\
	    MMoE & 0.7479$_{(+0.0023)}$ & \underline{0.6474}$_{(+0.0139)}$ & 0.9490$_{(-0.0001)}$ & 0.7980$_{(+0.0014)}$ & -&-&-&-\\
	    PLE & \underline{0.7485}$_{(+0.0029)}$ & 0.6464$_{(+0.0129)}$ & \underline{0.9495}$_{(+0.0004)}$ & 0.7983$_{(+0.0017)}$ & -&-&-&-\\
	    \specialrule{0em}{1pt}{1pt} \hline \specialrule{0em}{1pt}{1pt}
	    TAUG & 0.7491$_{(+0.0035)}$ & 0.6743$_{(+0.0408)}$ & 0.9498$_{(+0.0007)}$ & 0.8081$_{(+0.0115)}$ & 0.7485$_{(+0.0032)}$ & \textbf{0.7408}$_{(+0.0268)}$ & 0.9501$_{(+0.0020)}$ & \textbf{0.8335}$_{(+0.0038)}$\\
	    CrossDistil & \textbf{0.7494}$_{(+0.0038)}$ & \textbf{0.7411}$_{(+0.1076)}$ & \textbf{0.9513}$_{(+0.0022)}$ & \textbf{0.8341}$_{(+0.0375)}$ & \textbf{0.7487}$_{(+0.0034)}$ & 0.7403$_{(+0.0263)}$ & \textbf{0.9502}$_{(+0.0021)}$ & 0.8324$_{(+0.0027)}$\\
		\specialrule{0em}{1pt}{1pt}
		\bottomrule[1pt]
	\end{tabular}}
% 	\vspace{-5pt}
\end{table*}

\subsection{Evaluation Metrics}
We use two widely adopted metrics, i.e., AUC and Multi-AUC, for evaluation. AUC indicates the bipartite ranking (i.e., $\mathbf x_{+} \succ \mathbf x_{-}$) performance of the model. 
\begin{equation}
    \mbox{AUC} = \frac{1}{N^+N^-}\sum_{\mathbf x_{i}\in D^+}\sum_{\mathbf x_{j}\in D^-}(\mathcal{I}(p(\mathbf x_{i})>p(\mathbf x_{j})))
\end{equation}
% where $D^+$ (resp. $D^-$) is the collection of all positive (resp. negative) samples. 
where $p(\mathbf x)$ is the predicted probability of $\mathbf x$ being a positive sample and $\mathcal{I}(\cdot)$ is the indicator function. 
% Multi-AUC measures the multipartite ranking performance, i.e., $\mathbf x_{_{++}} \succ \mathbf x_{_{+-}} \succ \mathbf x_{_{-+}} \succ \mathbf x_{_{--}}$.

\paragraph{Multi-Class Area Under ROC Curve (Multi-AUC)} The vanilla formulation of AUC only measures the performance of bipartite ranking where a data point is labeled either as a positive sample or a negative one. However, we are also interested in multipartite ranking performance since samples have multiple classes with an order $\mathbf x_{_{++}} \succ \mathbf x_{_{+-}} \succ \mathbf x_{_{-+}} \succ \mathbf x_{_{--}}$ (for task $A$). Therefore, following~\cite{shan2018combined,shan2017optimizing}, we adopt Multi-AUC to evaluate multipartite ranking performance on test set. Note that we use the weighted version which considers the class imbalance problem~\cite{hand2001simple} and is defined as:
\begin{equation}
    \mbox{Multi-AUC} = \frac{2}{c(c-1)} \sum_{j=1}^c \sum_{k>j}^c p(j\cup k)\cdot AUC(k,j),
\end{equation}
where $c$ is the number of classes, $p()$ is the prevalence-weighting function as described in~\cite{ferri2009experimental}, $AUC(k,j)$ is the AUC score with class $k$ as the positive class and $j$ as the negative class.

\subsection{Baseline Methods}
 We choose the following MTL models with different shared network architectures for comparison: Shared-Bottom~\cite{caruana1997multitask}, Cross-Stitch~\cite{misra2016cross}, MMoE~\cite{ma2018modeling}, PLE~\cite{tang2020progressive}. 
 We use two variants of our method: TAUG incorporates augmented tasks on top of MTL models, and CrossDistil extends TAUG by conducting calibrated knowledge distillation. Despite that Both TAUG and CrossDistil could be implemented on most state-of-the-art MTL models, \emph{we choose the best competitor (i.e. PLE) as the backbone}.

\subsection{RQ1: Performance Comparison}

% Table~\ref{tab_result1} and ~\ref{tab_result2} shows the experiment results of our model versus other state-of-the-art MTL competitors on TikTok and Production dataset respectively. The bold value marks the best one in one column, while the underlined value corresponds to the best one among all the baselines. To show the improvements over non-MTL model, we also report results of Single-Model which uses a separate network for learning tasks on both datasets. To offer more insights, we also report performance of the proposed augmented ranking-based tasks as single tasks independent from regression-based tasks and other augmented tasks. As is shown in the tables, the Shared-Bottom model in most cases is the worst competitor among all baseline models presumably because of task conflicts or negative transfer problem. In contrast, the proposed CrossDistil achieves the best performance improvements over the single task model in terms of AUC and Multi-AUC. These results manifest that CrossDistil could indeed better leverage the knowledge from other tasks to improve the ranking ability of the main tasks. 

Table~\ref{tab_result2} and ~\ref{tab_result1} show the experiment results of our methods versus other competitors on WechatMoments and TikTok datasets respectively. The bold value marks the best one in one column, while the underlined value corresponds to the best one among all the baselines. To show improvements over the single-task counterpart, we report results of Single-Model which uses a separate network for learning each task. 
% To offer more insights, we also report performance of the proposed augmented ranking-based tasks as single task. 
As is shown in the tables, the proposed CrossDistil achieves the best performance improvements over Single-Model in terms of AUC and Multi-AUC~\footnote{For large-scale datasets in online advertisement, the improvements of AUC in the table is considerable because of its hardness.}. These results manifest that CrossDistil could indeed better leverage the knowledge from other tasks to improve both bipartite and multipartite ranking abilities on all tasks. Also, TAUG model alone, without calibrated KD, achieves better performance compared with the backbone model PLE, which validates the effectiveness of task augmentation.

% Second, we observe that  This is because the introduced ranking-based tasks could learn unconflicted fine-grained ranking and provide hints about what shall be learned and transferred in shared layers. CrossDistil even has better performance than TAUG, which verities the effectiveness of calibrated knowledge distillation.

\begin{table}[t]
\centering
\caption{Ablation analysis for Task A on TikTok dataset.
% 		\vspace{-5pt}
\label{tab_ab1}}
\scalebox{0.72}{\setlength{\tabcolsep}{6mm}{
\begin{tabular}{@{}l|c|c}
\toprule
% \multirow{2}{*}{Variants} & \multicolumn{2}{c}{Task A} & \multicolumn{2}{c}{Task B} \\ \cmidrule(l){2-5} 
Variants & AUC       & Multi-AUC         \\ \midrule
w/o AuxiliaryRank   & 0.7488 ${(-0.0006)}$      & 0.6510 ${(-0.0901)}$ \\
w/o Calibration     & 0.7478 ${(-0.0016)}$      & 0.7396 ${(-0.0015)}$ \\
w/o Correction      & 0.7486 ${(-0.0008)}$      & 0.7399 ${(-0.0012)}$ \\
KD (same task)          & 0.7489 ${(-0.0005)}$      & 0.6901 ${(-0.0510)}$ \\
KD (cross task)          & 0.7269 ${(-0.0225)}$      & 0.6120 ${(-0.1291)}$ \\
\midrule
Baseline & \textbf{0.7494} & \textbf{0.7411} \\ \bottomrule
\end{tabular}}}
% \vspace{-6pt}
\end{table}

\begin{table}[t]
\centering
\caption{Ablation analysis for Task B on TikTok dataset. \label{tab_ab2}}
% 		\vspace{-5pt}
\scalebox{0.72}{\setlength{\tabcolsep}{6mm}{
\begin{tabular}{@{}l|c|c}
\toprule
% \multirow{2}{*}{Variants} & \multicolumn{2}{c}{Task A} & \multicolumn{2}{c}{Task B} \\ \cmidrule(l){2-5} 
Variants & AUC       & Multi-AUC           \\ \midrule
w/o AuxiliaryRank       & 0.9501 ${(-0.0012)}$      & 0.8005 ${(-0.0336)}$       \\
w/o Calibration          & 0.9504 ${(-0.0009)}$      & 0.8312 ${(-0.0029)}$       \\
w/o Correction         & 0.9508 ${(-0.0005)}$      & 0.8310 ${(-0.0031)}$       \\
KD (same task)         & 0.9505 ${(-0.0008)}$      & 0.8014 ${(-0.0327)}$       \\
KD (cross task)         & 0.9184 ${(-0.0329)}$      & 0.7520 ${(-0.0821)}$       \\
\midrule
Baseline & \textbf{0.9513} & \textbf{0.8341}  \\ \bottomrule
\end{tabular}}}
% \vspace{-5pt}
\end{table}

\begin{figure*}[t]
\subfigure[Error correction margin $m$]{
\label{Fig.hp.1}
\includegraphics[width=0.23\textwidth]{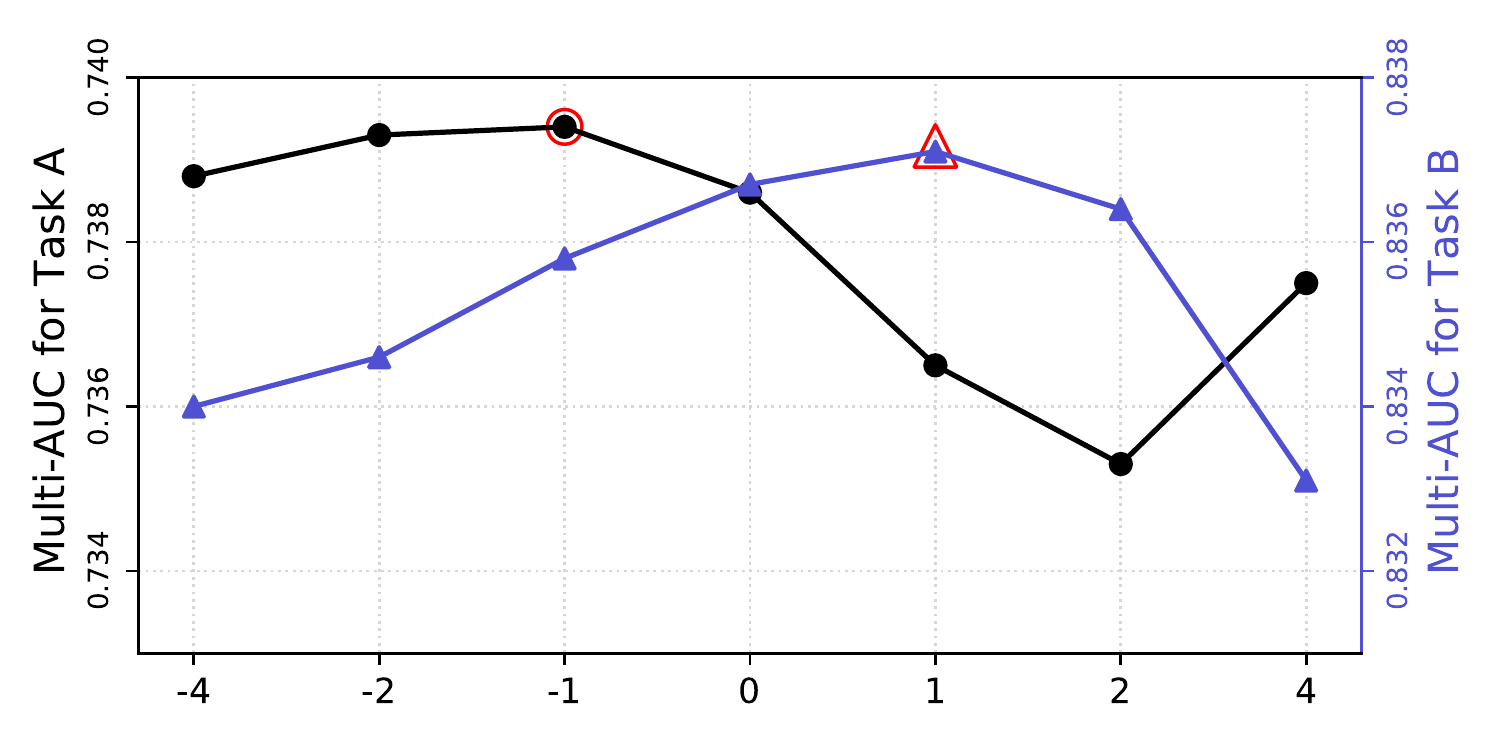}}
\subfigure[Coefficient $\beta_1$]{
\label{Fig.hp.2}
\includegraphics[width=0.23\textwidth]{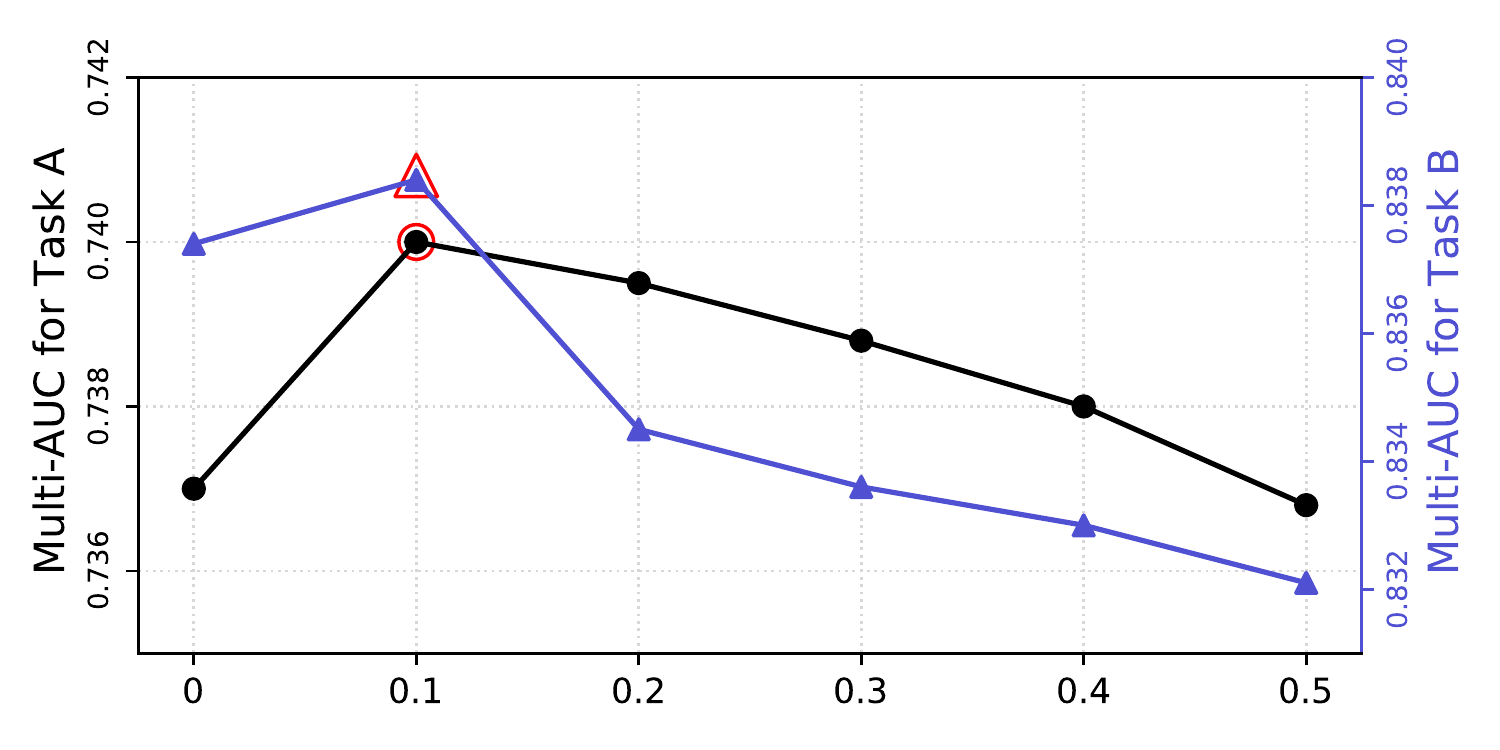}}
\subfigure[Coefficient $\beta_2$]{
\label{Fig.hp.3}
\includegraphics[width=0.23\textwidth]{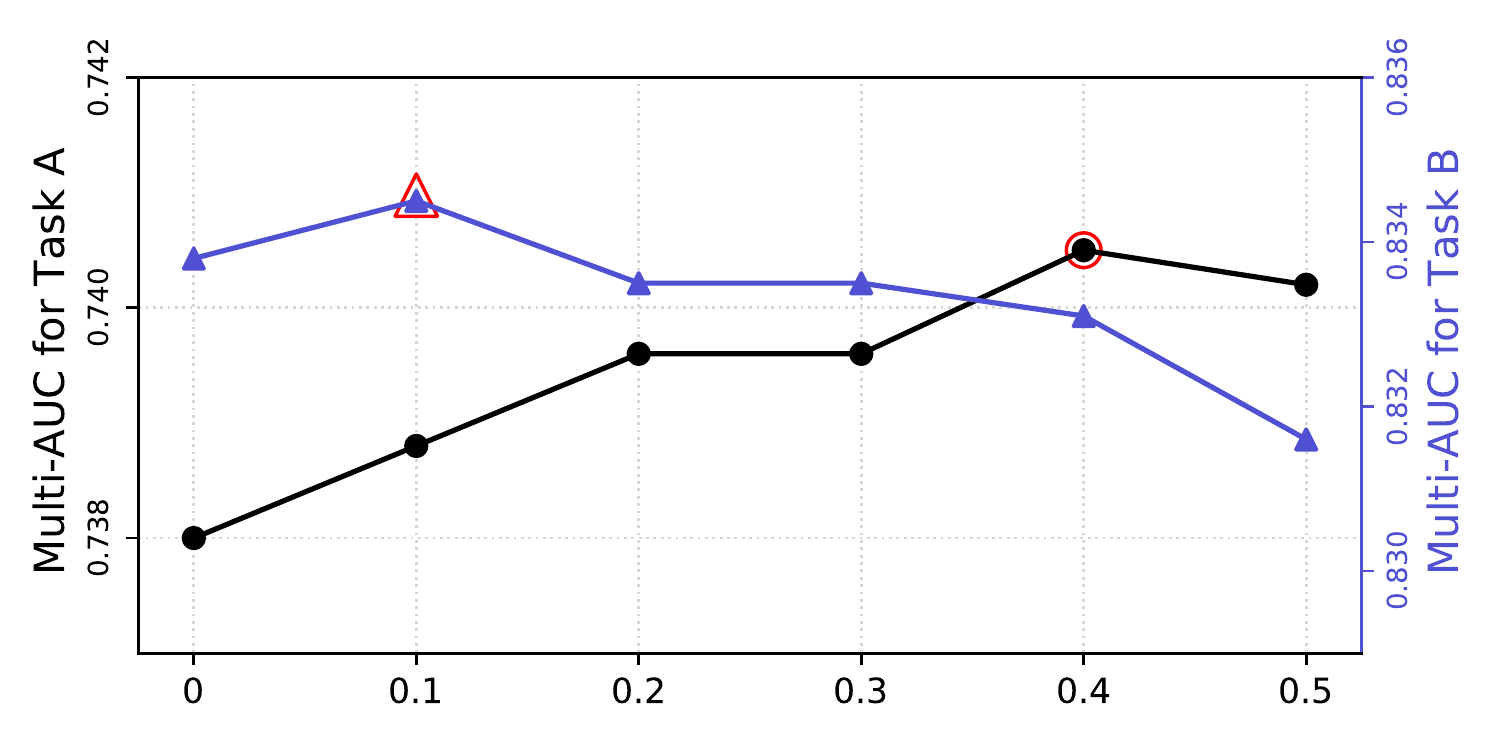}}
\subfigure[Distillation loss weight $\alpha$]{
\label{Fig.hp.4}
\includegraphics[width=0.23\textwidth]{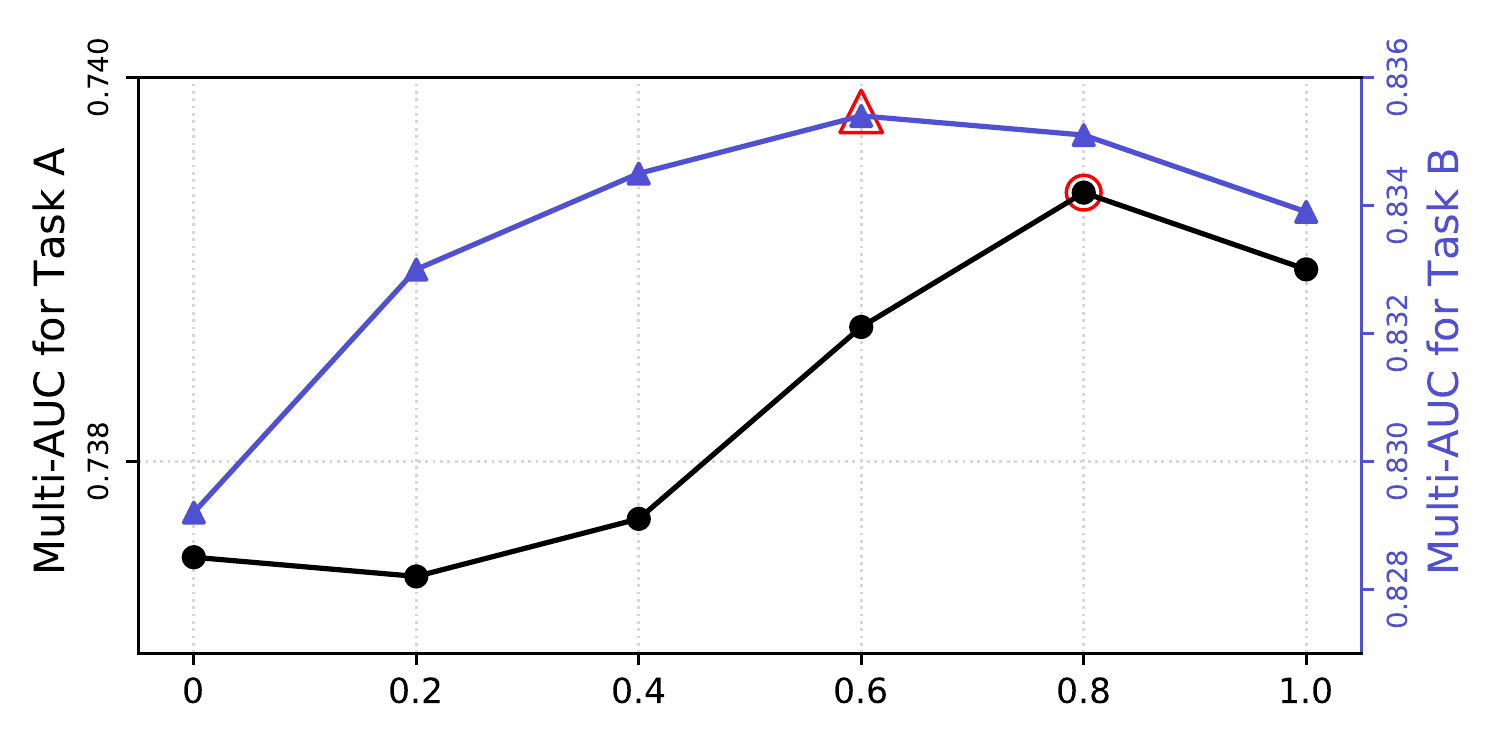}}
\caption{Multi-AUC performance on TikTok dataset for TaskA and Task B w.r.t. different hyper-parameters.}
% \vspace{-5pt}
\label{fig_hyper}
\end{figure*}

\begin{figure}[t]
% \vspace{-10pt}
\centering
\subfigure[Task A (Finish Watching).]{
\begin{minipage}[t]{0.45\linewidth}
\centering
\includegraphics[width=\textwidth]{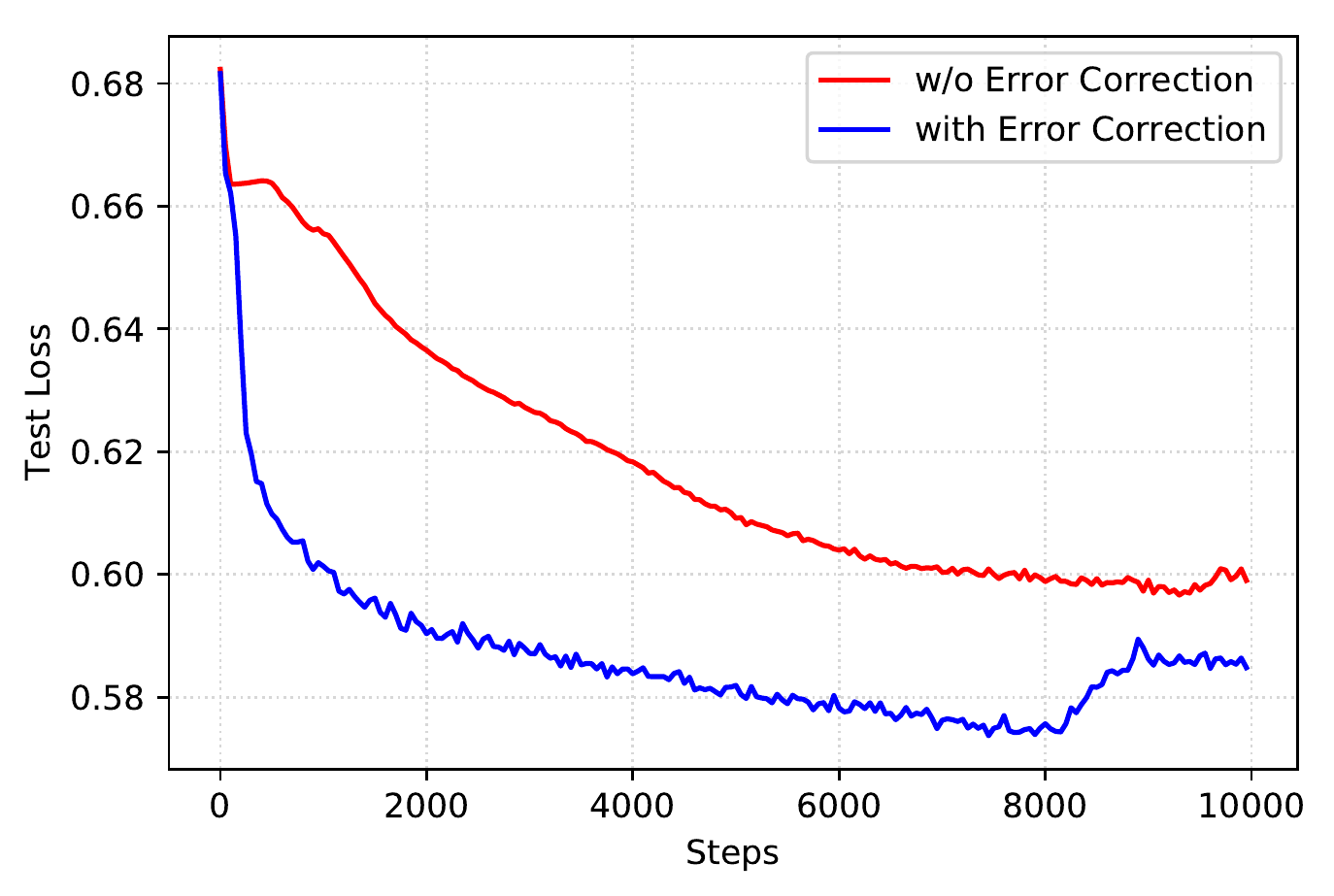}
%\caption{fig2}
\end{minipage}
}%
\subfigure[Task B (Like).]{
\begin{minipage}[t]{0.45\linewidth}
\centering
\includegraphics[width=\textwidth]{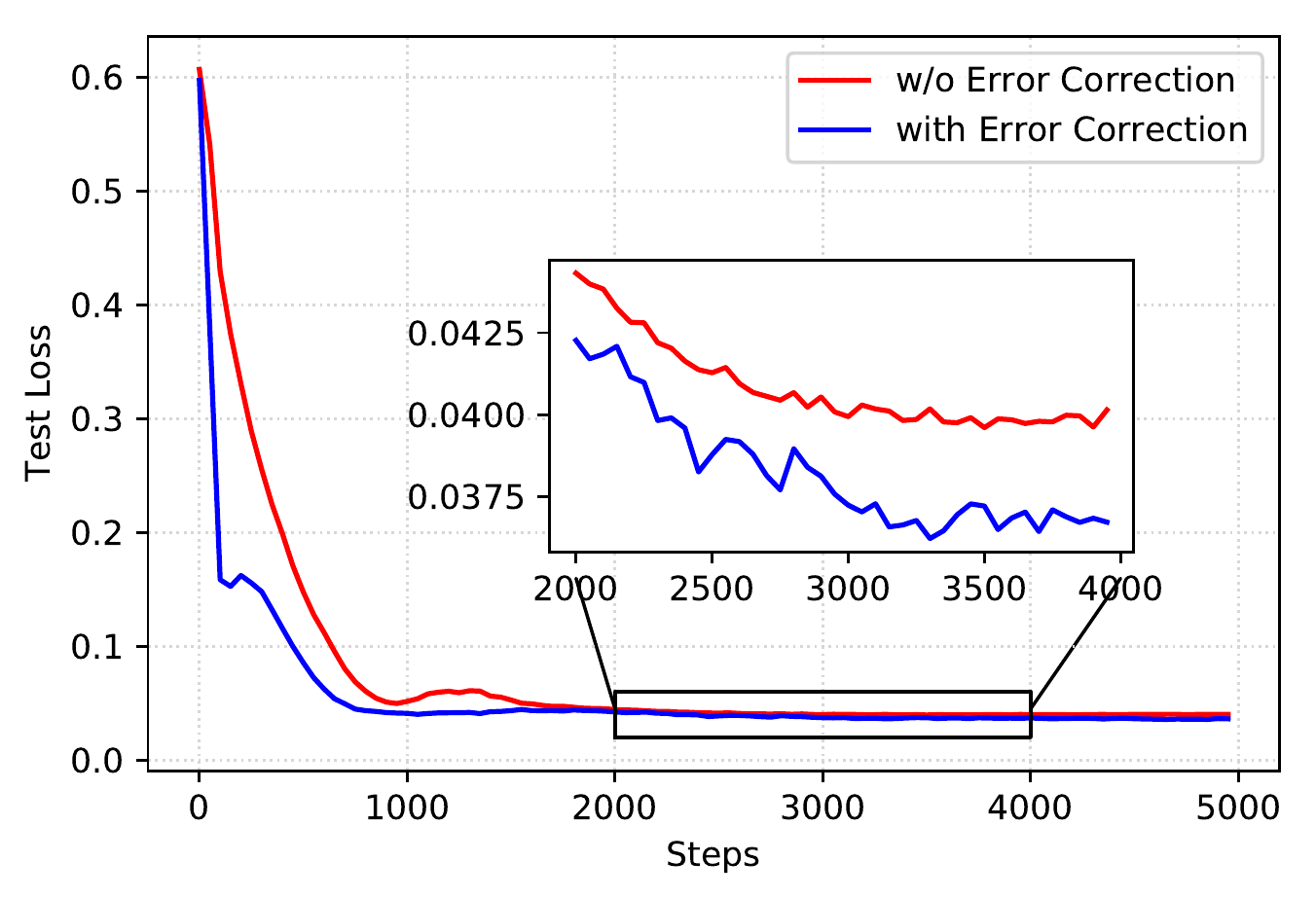}
%\caption{fig2}
\end{minipage}
}%
\centering
\caption{Learning curves of CrossDistil with and without error correction mechanism on TikTok dataset. \label{fig_curve}}
% \vspace{-15pt}
\end{figure}

\begin{figure}[t]
\centering
\subfigure[Task A (Finish Watching).]{
\begin{minipage}[t]{0.45\linewidth}
\centering
\includegraphics[width=\textwidth]{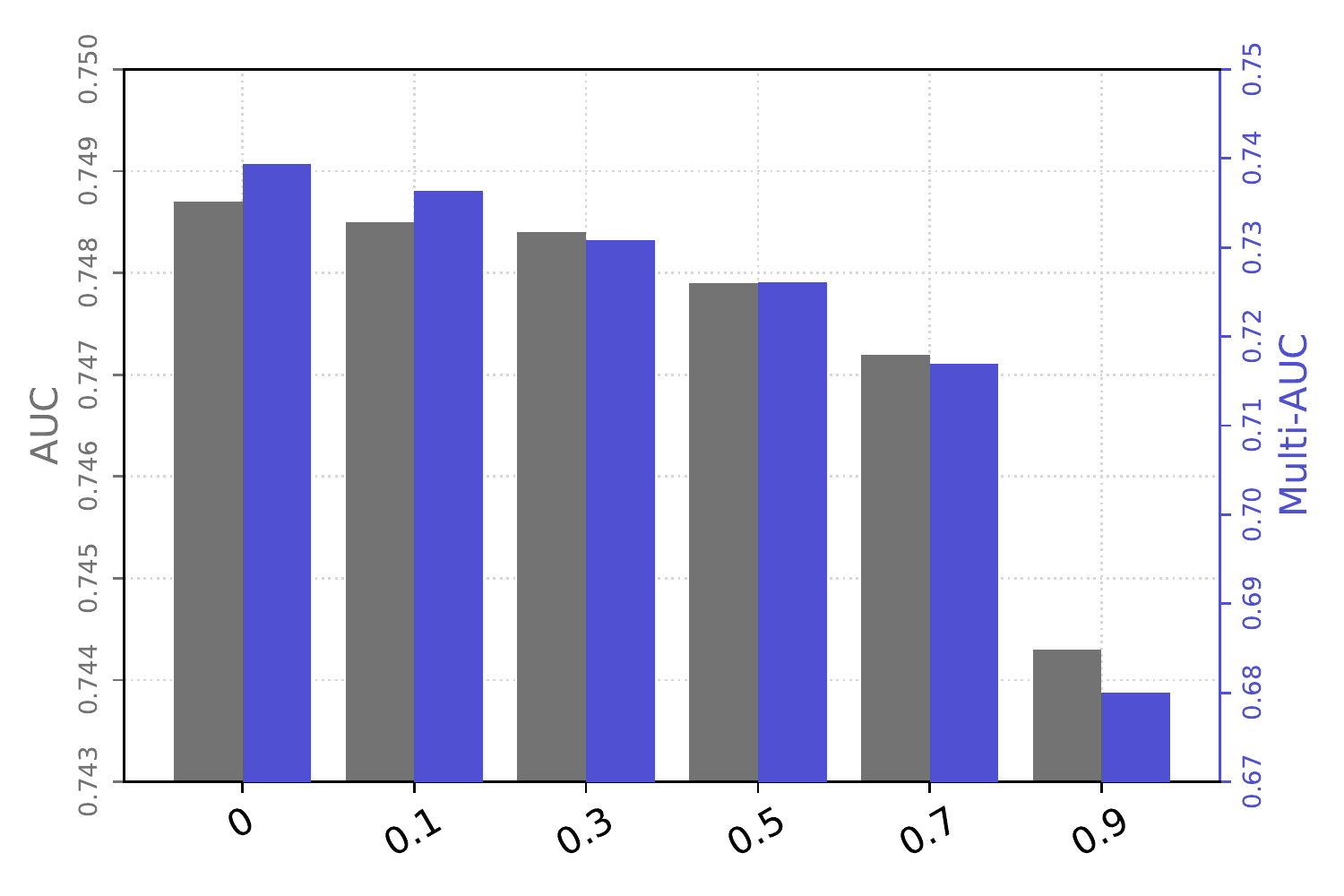}
%\caption{fig2}
\end{minipage}
}%
\subfigure[Task B (Like).]{
\begin{minipage}[t]{0.45\linewidth}
\centering
\includegraphics[width=\textwidth]{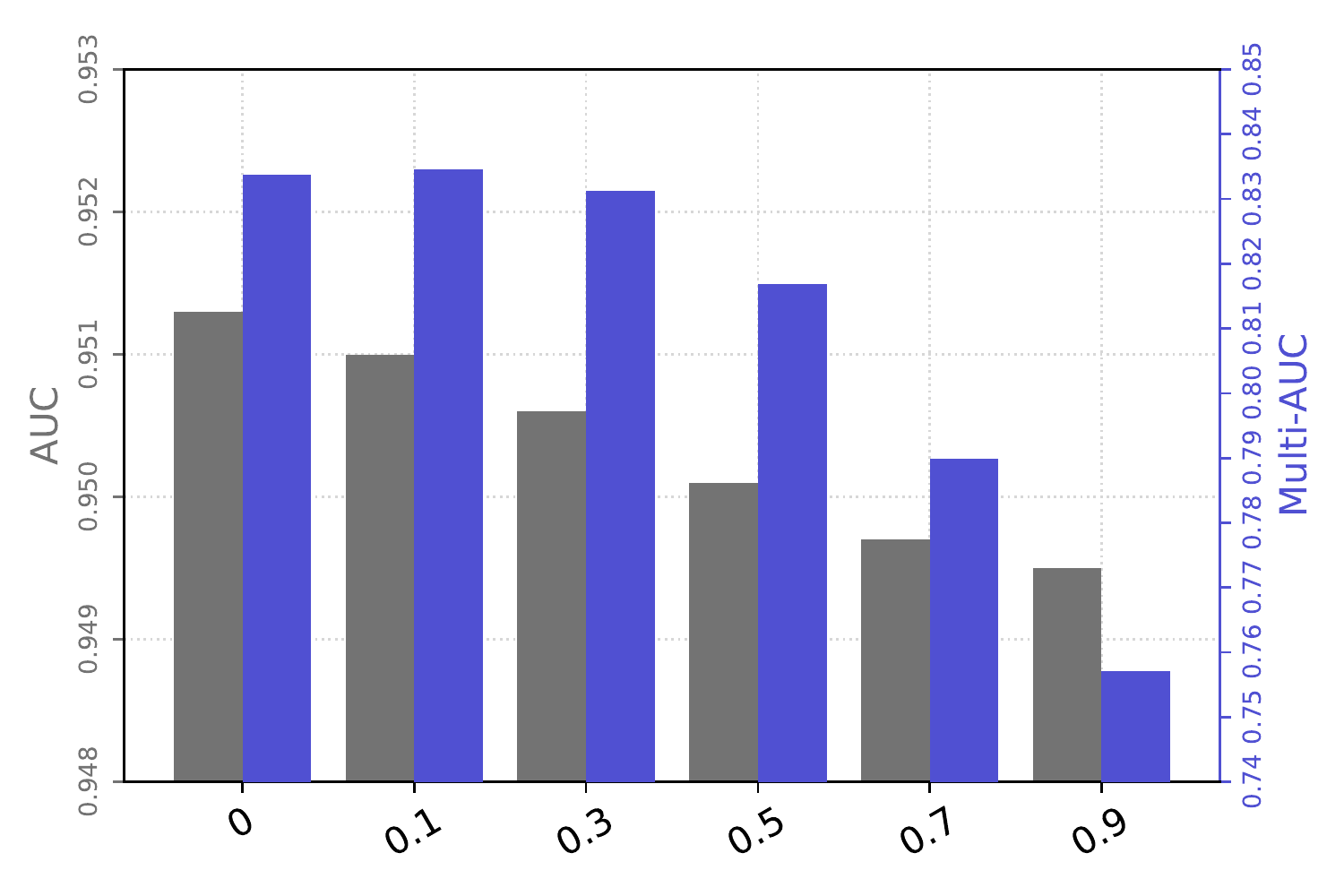}
%\caption{fig2}
\end{minipage}
}%
\centering
\caption{Impact of corrupted auxiliary ranking information on the student model performance for TikTok dataset. \label{fig_cor}}
% \vspace{-15pt}
\end{figure}

Besides, there are several observations in comparison tables. First, Single-Model on augmented ranking-based tasks (teacher) achieves better results in Multi-AUC compared with Single-Model on original regression-based task (student). It verifies that the proposed augmented tasks are capable of capturing task-specific fine-grained ranking information. Second, the student model exceeds the teacher model both in AUC and Multi-AUC performance in most cases, which is not strange since the student benefits from additional training signals that could act as label smoothing regularization and the teacher does not have such advantage. The same phenomenon is observed in many other works~\cite{yuan2020revisiting,tang2020understanding,zhang2020self}

\subsection{RQ2,3,4: Ablation Study}

% In order to investigate the effectiveness of some key components of CrossDistil and how these components contribute to the overall results, we design a series of ablation studies for CrossDistil. Four variants are considered to simplify CrossDistil in different ways: 1) removing the second and third term of teacher model loss function in Eqn~\eqref{eqn_apr}, 2) directly using teacher model outputs as soft labels by not optimizing calibration loss, 3) not correcting predictions of teacher model as in Eqn~\eqref{eqn_corr}, 4) replacing augmented ranking-based model with regression-based model for teacher network and using the vanilla knowledge distillation, which is similar with~\cite{zhou2018rocket}. Table~\ref{tab_ab1} and ~\ref{tab_ab2} show the results of AUC and Multi-AUC for these variants on TikTok dataset, and their relative performance drop compared with the baseline, i.e., the original CrossDistil. 

We design a series of ablation studies to investigate the effectiveness of some key components. Four variants are considered to simplify CrossDistil by: i) removing BPR losses for learning auxiliary ranking relations, ii) directly employing the teacher model outputs for knowledge distillation without any calibration, iii) not applying the error correction mechanism, vi) using regression-based teacher models that learn the same task as students and using the vanilla knowledge distillation that is similar with~\cite{zhou2018rocket}, v) directly using the predictions of another task for distillation. Table~\ref{tab_ab1} and ~\ref{tab_ab2} show the results for these variants on TikTok dataset and performance drops compared with the baseline (i.e. CrossDistil).

For the first variant, teacher loss function degrades to traditional BPR loss with no auxiliary ranking information. Such auxiliary ranking information that contains cross-task knowledge is a key factor for good performance in AUC and Multi-AUC. The second variant without calibration may produce unreliable soft labels and result in performance deterioration. Also, it is worth mentioning that the calibration process could significantly improve the performance of LogLoss, which is a widely used regression-based metric. Concretely, LogLoss reduces from $0.5832$ to $0.5703$ for task $A$, and $0.0623$ to $0.0337$ for task $B$ by using calibration. The results of the third variant indicate that the error correction mechanism can also bring up improvements for AUC and Multi-AUC. Another benefit of error correction is to accelerate model training, which will be further discussed. For the fourth variant, we can see that the proposed CrossDistil is better than the vanilla KD since it transfers fine-grained ranking knowledge across tasks. For the last variant, directly conducting KD could cause performance drop because of the ranking conflicts of tasks.
% Such performance enhancement could be explained by two reasons: 1) the auxiliary ranking-based teacher model explicitly learns fine-grained order information and implicitly shares such knowledge in the bottom layer, while the teacher model for traditional KD could not learn such information; 2) the calibrated knowledge distillation explicitly transfers fine-grained ranking knowledge through calibrated soft labels, where traditional KD fails to do so.

\paragraph{RQ3: Does Error Correction Mechanism Help to Accelerate Convergence and Enhance Knowledge Quality?} To answer this question, we plot the learning curves of test loss with (blue line) and without (red line) error correction in Fig.~\ref{fig_curve}. As we can see, for both tasks, the test loss of CrossDistil with error correction significantly goes down faster at the beginning of the training process when the teacher is not well-trained. Plus, at later stage of training when the teacher becomes well-trained, the test loss of CrossDistil with error correction slowly keeps going down and achieves better optimal results compared with the variant, indicating that the proposed error correction mechanism could indeed help to improve knowledge quality. 

% On the other hand, compared with the variant of CrossDistil without error correction mechanism, our model has better optimal test loss, which could attribute to its ability to handling label conflicts in the later stage of training.

\paragraph{RQ4: Does the Student Model Really Benefit from Auxiliary Ranking Knowledge from Other Tasks?} To answer this question, we conduct the following experiment: For a target task $A$, we randomly choose a certain ratio of positive samples of task $B$, and then exchange their task $B$'s label with the same number of randomly selected negative samples, to create a corrupted training set. Note that such data corruption process only has negative effects on the reliability of the auxiliary ranking information, so that we can investigate its impact on the student model's performance. Figure~\ref{fig_cor} shows the results of performance change when increasing the ratio from $10\%$ to $90\%$. The results indicate that flawed auxiliary information has considerable negative effects on the overall performance, which again verifies CrossDistil could effectively transfer knowledge across tasks.

\subsection{RQ5: Hyper-parameter Study}
% This subsection studies the performance variation of CrossDistil w.r.t some key hyper-parameters (i.e. error correction margin $m$, auxiliary ranking loss coefficient $\beta_1$ and $\beta_2$, distillation loss weight $\alpha$), some of which are firstly proposed by this work. Figure~\ref{Fig.hp.1} shows the performance of Multi-AUC with error correction margin ranges from $-4$ to $4$. As we can see, the model performance first increases and then decreases. This is because very small $m$ is equivalent to not conducting error correction, and vert large $m$ makes the soft labels degrade to ground truth label. In Fig.~\ref{Fig.hp.2} and Fig.~\ref{Fig.hp.3}, we change two auxiliary ranking loss coefficients $\beta_1$ and $\beta_2$ from $0$ to $0.5$. The model performance also shows a trend from decline to rise as $\beta$ increases. This indicates a proper setting for $\beta$ can help to capture the correct underlying fine-grained ranking information. Figure.~\ref{Fig.hp.4} shows the performance change w.r.t variable $\alpha$. As a result, a proper $\alpha$ from $0$ to $1$ can bring the best performance, which is reasonable since distillation loss plays the role of label smoothing regularization to provide guidance for student model.

This subsection studies the performance variation of CrossDistil w.r.t. some key hyper-parameters (i.e. error correction margin $m$, auxiliary ranking loss coefficient $\beta_1$ and $\beta_2$, distillation loss weight $\alpha$). Figure~\ref{Fig.hp.1} shows the Multi-AUC performance with error correction margin ranges from $-4$ to $4$. As we can see, the model performance first increases and then decreases. This is because extremely small $m$ is equivalent to not conducting error correction, while extremely large $m$ makes the soft labels degrade to hard labels. The results in Fig.~\ref{Fig.hp.2} and Fig.~\ref{Fig.hp.3} indicate a proper setting for $\beta$ can help to capture the correct underlying fine-grained ranking information. The results in Fig.~\ref{Fig.hp.4} reveal that a proper $\alpha$ from $0$ to $1$ can bring the best performance, which is reasonable since the distillation loss plays the role of label smoothing regularization and could not replace hard labels.

\section{Conclusion}

In this paper, we propose a cross-task knowledge distillation framework for multi-task recommendation. First, augmented ranking-based tasks are designed to capture fine-grained ranking knowledge, which could avoid conflicted information to alleviate negative transfer problem and prepare for subsequent knowledge distillation. Second, calibrated knowledge distillation is adopted to transfer knowledge from augmented tasks (teacher) to original tasks (student). Third, an additional error correction method is proposed to speed up the convergence and improve knowledge quality in the synchronous training process.

CrossDistil could be incorporated in most off-the-shelf multi-task learning models, and is easy to be extended or modified for industrial applications such as online advertising. The core idea of CrossDistil could inspire a new paradigm for solving domain-specific task conflict problem and enhancing knowledge transfer in broader areas of data mining and machine learning.

% Use \bibliography{yourbibfile} instead or the References section will not appear in your paper

\section{Acknowledgments}
This work was supported by the National Key R\&D Program of China [2020YFB1707903]; the National Natural Science Foundation of China [61872238, 61972254], Shanghai Municipal Science and Technology Major Project [2021SHZDZX0102], the Tencent Marketing Solution Rhino-Bird Focused Research Program [FR202001], the CCF-Tencent Open Fund [RAGR20200105], and the Huawei Cloud [TC20201127009]. Xiaofeng Gao is the corresponding author.

\bibliography{aaai22}

\end{document}